\documentclass[prl,twocolumn,superscriptaddress,showpacs]{revtex4}

\usepackage{amsmath,amssymb}
\usepackage{graphicx}

\allowdisplaybreaks

\preprint{IFT-UAM/CSIC-07-55}
\preprint{YITP-SB-07-34}

\begin{document}

\title{Radiography of the Earth's Core and Mantle with Atmospheric
  Neutrinos}

\author{M.C.~Gonzalez-Garcia}
\affiliation{%
  Instituci\'o Catalana de Recerca i Estudis Avan\c{c}ats (ICREA),
  Departament d'Estructura i Constituents de la Mat\`eria, 647
  Diagonal, E-08028 Barcelona, Spain}
\affiliation{%
  C.N.~Yang Institute for Theoretical Physics, SUNY at Stony Brook,
  Stony Brook, NY 11794-3840, USA}

\author{Francis Halzen}
\affiliation{%
  Department of Physics, University of Wisconsin, Madison, WI 53706,
  USA}

\author{Michele Maltoni}
\affiliation{%
  Departamento de F\'isica Te\'orica \& Instituto de F\'isica
  Te\'orica UAM/CSIC, Facultad de Ciencias C-XI, Universidad
  Aut\'onoma de Madrid, Cantoblanco, E-28049 Madrid, Spain}

\author{Hiroyuki K.M.~Tanaka}
\affiliation{%
  Earthquake Research Institute, University of Tokyo, 113-0032 Tokyo,
  Japan}
\affiliation{%
  Atomic Physics Laboratory, RIKEN, 351-0198 Saitama, Japan}

\begin{abstract}
    A measurement of the absorption of neutrinos with energies in
    excess of 10~TeV when traversing the Earth is capable of revealing
    its density distribution. Unfortunately, the existence of beams
    with sufficient luminosity for the task has been ruled out by the
    AMANDA South Pole neutrino telescope. In this letter we point out
    that, with the advent of second-generation kilometer-scale
    neutrino detectors, the idea of studying the internal structure of
    the Earth may be revived using atmospheric neutrinos instead.
\end{abstract}


\maketitle

The density profile and the shape of the Earth's core and mantle and
their boundary (CMB) determine its geodynamo as well as the feeding
mechanism of hotspots at the surface~\cite{Loper}. Knowledge of the
CMB is derived from body-wave and free oscillation studies. The
information, while more precise than what we can realistically expect
from neutrino radiography in the near future, cannot reduce
ambiguities in our present model of the CMB associated with the fact
that arrays of seismometers only provide regional information, and
that free-oscillation data only reveal one dimensional structure.  The
trade off among density, temperature, and chemical structure for body
wave studies increases the uncertainty of the value for the density.
For these reasons aspects of the global structure of the CMB region
require confirmation. The study presented in this paper indicates that
present neutrino detectors have to be operated for 10 years to locate
the CMB.  IceCube will establish the averaged core and mantle density
as a function of longitude thus providing the first independent global
survey of the CMB region.  We anticipate however that more precise
global information on the CMB region will be obtained by longer
observation periods or by future large scale neutrino detectors. Early
studies of the possibility of doing neutrino tomography date back more
than 25 years~\cite{tomo1-7}. These proposed studying the passage of
cosmic beams of high energy (HE) neutrinos through the Earth to
diagnose its density. Alternatively, it was suggested to use
accelerator beams~\cite{charpak} and to study the propagation effects
through matter of oscillating neutrinos; for a review see
Ref.~\cite{winter}.

The idea of neutrino tomography is straightforward: the Earth becomes
opaque to neutrinos whose energy exceeds $\sim 10$~TeV. The diameter
of the Earth represents one absorption length for a neutrino with an
energy $\sim$ 25~TeV. Such neutrinos are produced in collisions of
cosmic rays with nuclei in the Earth's atmosphere but, because of the
steeply falling energy spectrum of $\sim E_\nu^{-3.7}$ of the
atmospheric neutrino flux, such events are rare. The hope was that
beams of cosmic neutrinos, likely to be associated with the sources of
the cosmic rays which reach energies of $10^8$~TeV, would provide a
plentiful source of neutrinos in the appropriate energy range. These
would be detected by HE neutrino telescopes under development at the
time.  The cosmic beams would perform radiography of the Earth's
interior as it moves relative to the comic source. In light of the
recent development of successful and affordable technologies to build
very large neutrino telescopes, we revisit the
proposal~\cite{halzini}.

Neutrino telescopes detect the Cherenkov radiation from secondary
particles produced in the interactions of HE neutrinos in deep water
or ice. At the higher energies the neutrino cross section grows and
secondary muons travel up to tens of kilometers to reach the detector
from interactions outside the instrumented volume~\cite{reviews}. The
construction of kilometer-scale instruments such as IceCube at the
South Pole and the future KM3NeT detector in the Mediterranean, have
been made possible by development efforts that resulted in the
commissioning of prototypes that are two orders of magnitude smaller,
AMANDA and ANTARES~\cite{antares}. Their successful technologies have,
in turn, relied on pioneering efforts by the DUMAND~\cite{dumand} and
Baikal~\cite{baikal}, as well as the Macro and Super-Kamiokande
collaborations~\cite{reviewslow}.  IceCube~\cite{ice3} is under
construction and taking data with a partial array of 1320 ten inch
photomultipliers positioned between 1500 and 2500 meter, deployed as
beads on 22 strings below the geographic South Pole. Its effective
area already exceeds that of its predecessor AMANDA by $\sim$ one
order of magnitude. The detector will grow by other $14{\sim}18$
strings in 2007-08 to be completed in 2011 with 80 strings.

AMANDA has observed neutrinos with energies as high as
${\sim}100$~TeV, at a rate consistent with the flux of atmospheric
neutrinos (ATM-$\nu$'s) extrapolated from lower energy measurements.
It thus establishes limits on any additional flux of cosmic neutrinos
in the energy range of interest for Earth tomography. These now reach
below $E_\nu^2 \, dN / dE_\nu < 10^{-11}~\text{TeV} \, \text{cm}^{-2}
\, \text{s}^{-1} \, \text{sr}^{-1}$ for a diffuse flux~\cite{hill} and
$E_\nu^2 \, dN / dE_\nu < 10^{-10}~\text{TeV} \, \text{cm}^{-2} \,
\text{s}^{-1} \, \text{sr}^{-1}$~\cite{point}. From these results
Standard Model physics is sufficient to establish that the event rates
from cosmic beams per year in a future kilometer-scale detector are
limited to $\sim 10$ events from any particular source in the sky and
less than $\sim 100$ from the aggregate of sources. Needless to say
that the statistics is already uncomfortably small for a beam to be
exploited for Earth tomography.

Our main observation is that, with the growth of the detectors, the
opportunity arises to exploit the ATM-$\nu$'s that represent the
background in the search for cosmic sources, as a beam for studying
the Earth. The key point is that, the statistics for ATM-$\nu$'s in
the $10$ to $100$~TeV energy range, is superior to those expected from
any cosmic sources detected in the future within the upper limits
already established by AMANDA observations. Viewed from the South
Pole, a uniform flux of ATM-$\nu$'s reaches the detector from the
northern half of the sky; it will be modified in the 10~TeV energy
region by its passage through the Earth. For instance, neutrinos from
vertical to $\sim 30$~degrees have penetrated the core of the Earth
before detection, whereas the ones detected at larger angles have
traversed the mantle only. Establishing direct evidence for the
transition from mantle to core will here be used as a benchmark to
evaluate the technique. 

We will conclude that IceCube can directly observe the core-mantle
transition at the $5 \sigma$ level in 10 years. This evaluation is
based on modeling of the HE ATM-$\nu$'s that is, at present, still
subject to uncertainties. We however establish that, under
conservative assumptions, the transition can be observed at the $3
\sigma$ level or above.

We use the semianalytical calculation of IceCube event rates described
in Ref.~\cite{GHM}. In brief, the expected number of $\nu_\mu$-induced
events (events arising from $\bar\nu_\mu$ interactions can be
evaluated similarly) in an exposure time $T$ is:
\begin{multline}
    \label{eq:numuevents}
    N^{\nu_\mu}_\text{ev}
    = T \int^{1}_{-1} d\cos\theta \,
    \int_0^\infty dl'_\text{min} \,
    \int^\infty_{l'_\text{min}} dl \,
    \int_{E_\mu^\text{fin,min}}^\infty dE_\mu^\text{fin}
    \\
    \int_{E_\mu^\text{fin}}^\infty dE_\mu^0 \,
    \int_{E_\mu^0}^\infty dE_\nu
    \frac{d^2\phi_{\nu_\mu}}{dE_\nu \, d\cos\theta}(E_\nu,\cos\theta)
    \\
    \frac{d\sigma^\mu_{CC}}{dE_\mu^0}(E_\nu,E_\mu^0) \, n_T \,
    F(E_\mu^0, E_\mu^\text{fin}, l) \, A^0_\text{eff}
\end{multline}
where $d^2\phi_{\nu_\mu} / (dE_\nu \, d\cos\theta)$ is the
differential $\nu_\mu$ flux in the vicinity of the detector after
propagation in the Earth; more below.  We use as input the neutrino
fluxes from Honda~\cite{honda} extrapolated to match the fluxes from
Volkova~\cite{volkova} at higher energies.  At the relevant energies,
prompt $\nu_\mu$'s from charm decay are important. We introduce them
according to the recombination quark parton model~\cite{rqpm} but
consider alternative estimates also. 

\begin{figure}
    \includegraphics[width=0.99\columnwidth]{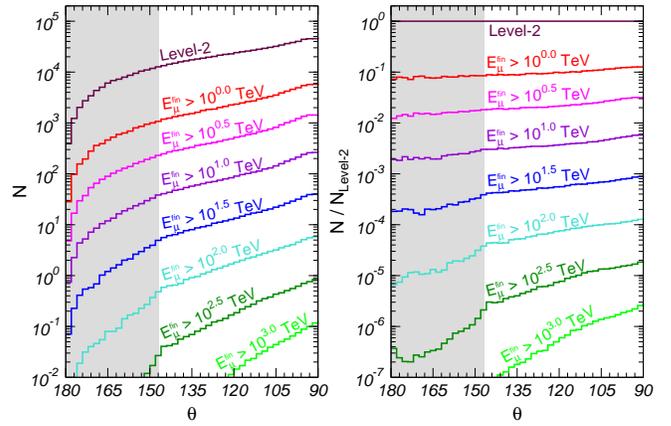}
    \caption{\label{fig:dist}%
      (a) Expected zenith angle distribution of ATM $\nu_\mu$ induced
      events in IceCube for different energy thresholds
      $E_\mu^\text{fin,min}$ for the PREM.  $\theta$ is the neutrino
      angle (which at these energies is collinear with the detected
      muon) as measured from the vertical direction (upgoing-$\nu$
      corresponding to $\theta = 180$).  (b) Ratio of the zenith angle
      distribution of ATM $\nu_\mu$ induced events in IceCube for
      different energy thresholds $E_\mu^\text{fin,min}$ over the
      corresponding one for L2 cuts only. The shadow areas cover the
      angular size of the Earth core.}
\end{figure}

$d\sigma^\mu_\text{CC} / dE_\mu^0(E_\nu,E_\mu^0)$ is the differential
interaction cross section producing a muon of energy $E_\mu^0$, After
production the muon ranges out in the rock and in the ice surrounding
the detector and loses energy to ionization, bremsstrahlung, $e^+e^-$
pair production and nuclear interactions. This is encoded in
$F(E_\mu^0, E_\mu^\text{fin}, l)$~\cite{ls} which represents the
probability that a muon produced with energy $E_\mu^0$ reaches the
detector with energy $E_\mu^\text{fin}$ after traveling a distance
$l$. $n_T$ is the number density of nucleons in the matter 
surrounding the detector.

The details of the detector are encoded in the effective area
$A^0_\text{eff}$ for which we use the parametrization in
Ref.~\cite{GHM} describing the response of IceCube after the
background rejection quality cuts referred to as ``Level-2" (L2) cuts
in Ref.~\cite{ice3}. $l_\text{min}=300$~m is the minimum muon track
length required for the event to be detected.  Effectively
$A^0_\text{eff}$ vanishes for $E_\mu^\text{fin} \lesssim 100$~GeV. 

In order to obtain $d^2\phi_{\nu_\mu} / (dE_\nu \, d\cos\theta)$, one
must account for the simultaneous effects of oscillations and
inelastic interactions with the Earth matter which lead to the
attenuation of the neutrino flux which are different for $\nu_\tau$'s
and $\nu_\mu$'s.~\cite{hs}. This can be achieved by solving a set of
coupled evolution equations for the neutrino flux density matrix and
for the muon and tau fluxes~\cite{GHM}. In practice, due to the
steepness of the ATM-$\nu$ spectra and the small value of the relevant
$\Delta m^2_{atm} / E_\nu$, $d^2\phi_{\nu_\mu} / (dE_\nu \,
d\cos\theta)$ can be obtained from:
\begin{multline}
    \label{eq:fluxapp}
    \frac{d^2\phi_{\nu_\mu}}{dE_\nu \, d\cos\theta}(E_\nu, \theta, L) =
    \frac{d^2\phi_{\nu_\mu}^0}{dE_\nu \, d\cos\theta}(E_\nu,\theta)
    \\
    P_{\mu\mu}(E_\nu, L) \,
    \exp\lbrace -X(\theta)[\sigma_\text{NC}(E_\nu)
    + \sigma_\text{CC}^\alpha(E_\nu)] \rbrace \,,
\end{multline}
where $P_{\mu\mu}(E_\nu, L = 2R \left| \cos\theta \right|)$ is the
oscillation probability. For $E_\nu \gtrsim 1$~TeV,
$P_{\mu\mu}\simeq1$. $X(\theta)$ is the column density of the Earth,
and $R$ its radius:
\begin{equation}
    X(\theta) = N_A \, \int_0^{L=2R \left| \cos\theta \right|}
    \hspace{-12mm}
    \rho_\text{E}(\sqrt{R^2+z^2+2 R z \cos\theta}) \, dz \,.
\end{equation}
$N_A$ is the Avogadro number, and $\rho_\text{E}(r)$ is the Earth
matter density assumed to be spherically symmetric.

Equation~\eqref{eq:fluxapp} embodies the physics that makes Earth
tomography with HE neutrinos possible.  At sufficiently high energies,
$E_\nu \gtrsim 10$~TeV, the attenuation factor $\exp\lbrace
-X(\theta)[\sigma_\text{NC}(E_\nu) + \sigma_\text{CC}^\alpha(E_\nu)]
\rbrace$ becomes relevant. Thus measuring $N^{\nu_\mu}_\text{ev}$ one
can get information on $\rho_\text{E}(r)$.

In the left panel of Fig.~\ref{fig:dist} we show the expected zenith
angle distribution of atmospheric $\nu_\mu$-induced events in IceCube
for different $E_\mu^\text{fin,min}$ energy threshold as obtained
using the Earth matter density profile of the Preliminary Reference
Earth Model (PREM)~\cite{PREM}.  In the PREM the Earth consists of a
mantle extending to radial distance $r\sim 3000$ km below the Earth
surface and a core under it with a sharp core-mantle transition in
density of about a factor 2.  Thus neutrinos arriving with $\theta
\gtrsim 147$ degrees ($\cos\theta \lesssim -0.84$) will cross the core
in their way to the detector.

In the figure one notices, at sufficiently high energies, a reduction
of the number of events for trajectories which cross the core
resulting in a ``kink'' in the angular distribution around $\theta
\gtrsim 147$.  This feature is more clearly illustrated in the right
panel of Fig.~\ref{fig:dist} where we plot the ratio of the zenith
angle distribution of events with energies above
$E_\mu^\text{fin,min}$ divided by the number of events with no
additional energy cut, which effectively corresponds to events with a
threshold energy $E_\mu^\text{fin,L2} \sim 100$~GeV.

Figure~\ref{fig:dist} illustrates the potential of doing Earth
tomography with the IceCube ATM-$\nu$ samples. However one must 
realize that the angular dependence in the ratio shown in the right
panel of Fig.~\ref{fig:dist} is not only due to Earth attenuation
factor: there is an additional, Earth-independent, contribution from
the variation of the zenith angle distribution of the fluxes with
$E_\nu$ which does not cancel out in the ratio of events at different
energies.  In principle this effect could be removed by comparing the
ratio of upgoing ($\theta > 90$ degrees) and downgoing events ($\theta
< 90$ degrees). In practice, the overwhelming atmospheric muon
background makes the measurement of downgoing $\nu_\mu$ events
impossible at these energies.

\begin{figure}
    \includegraphics[width=0.99\columnwidth]{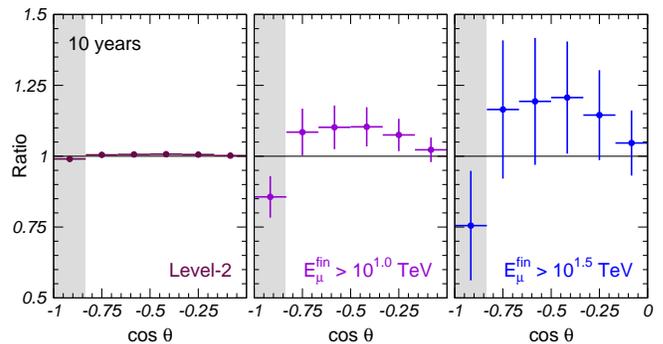}
    \caption{\label{fig:ratios}%
      Ratio of zenith angle distribution of expected events for the
      PREM over the expectations with an homogeneous Earth matter
      distribution for different values of the energy threshold of the
      events. The error bars in the figure show the expected
      statistical error in 10 years of IceCube.}
\end{figure}

In order to quantify the sensitivity of IceCube to the Earth density
profile we study the ratio of observed events above a given energy
threshold to the one expected for an Earth of equal mass as ours but
with an homogeneous matter distribution, $\rho_\text{hom} = 3
M_\text{Earth}/(4 \pi R^3)$:
\begin{equation}
    R = \frac
    {N_\mu (E_\mu^\text{fin} > E_\mu^\text{fin,min},\, \cos\theta,\, \rho_\text{PREM})}
    {N_\mu (E_\mu^\text{fin} > E_\mu^\text{fin,min},\, \cos\theta,\, \rho_\text{hom})}
\end{equation}
In Fig.~\ref{fig:ratios} we show this ratio obtained by integrating
the events in the numerator and denominator in 6 angular bins in
$\cos\theta$, and for three values of the threshold energy:
$E_\mu^\text{fin,L2}\sim$ 100~GeV, 10~TeV, and 32~TeV. In this plot,
trajectories crossing the core are contained in the most vertical bin.
In the figure we also show the expected statistical uncertainty
$\sigma_{\text{stat},i}$, computed from the expected number of events
in each angular bin in the PREM in 10 years of IceCube (see
Table~\ref{tab:nevents}).

\begin{table}
    \begin{tabular}{|c|c|c|c|}
        \hline 
        & \multicolumn{3}{c|} {$E_\mu^\text{fin,min}$} \\
        \hline%
       $[\cos\theta]$ & {$E_\mu^\text{fin,L2}$} & 10 TeV & 32 TeV \\ 
        \hline
        $[-1.00,-0.83]$  & 108320     & 254     & 27   \\
        $[-0.83,-0.67]$  & 115224     & 359     & 49   \\
        $[-0.67,-0.50]$  & 123524     & 429     & 62   \\
        $[-0.50,-0.33]$  & 137676     & 537     & 82   \\
        $[-0.33,-0.17]$  & 162500     & 736     &111   \\
        $[-0.17, 0.00]$  & 205500     & 1132    &169   \\
        \hline
    \end{tabular}
    \caption{\label{tab:nevents}%
      Number of expected atmospheric $\nu_\mu$-induced muon events in
      10 years of IceCube operation in the different angular bins and
      energy thresholds for the PREM.} 
\end{table}

As expected, events with low energy threshold have no sensitivity to
the Earth density and consequently the ratio for
$E_\mu^\text{fin,L2}\sim 100$~GeV is practically constant and equal to
1. As $E_\mu^\text{fin,min}$ increases the ratio becomes increasingly
different from 1, reflecting the fact that the effect of the Earth
matter profile becomes more evident.  The strategy is then obvious.
One uses the measured zenith angular distribution of the L2 event
sample as normalization to obtain the expectations for a constant
density Earth at higher energies, $N_\mu (E_\mu^\text{fin} >
E_\mu^\text{fin,min},\, \cos\theta,\, \rho_{\hom})$.  By comparing the
expectations with observation, one can quantify the sensitivity to the
Earth matter profile.

After normalizing to the observed L2 distribution residual theoretical
uncertainties remain associated with the predicted zenith angle
distribution.  They include systematic effects in calibration,
theoretical errors in the energy-angle dependence in the atmospheric
fluxes due to the uncertainties in the $K/\pi$ ratio as well as in the
contribution from charm (which are expected to be the largest at the
relevant energies), and the uncertainties in the neutrino interaction
cross sections.  Over the limited energy range relevant here, we
conservatively account for those by introducing three systematic
errors in the analysis: an overall normalization error of 20\%, and
angular \emph{tilt} uncertainty of 5\% between horizontal and vertical
events and an additional $\sigma_{\text{sys},i} = 1$\% uncertainty due
to uncorrelated systematics for each angular bin. With this we
construct two simple $\chi^2$ functions as
\begin{align}
    \label{eq:chi2nh}
    \chi^2_\text{nh}
    &= \mathop{\mathrm{Min}}_{\substack{
	\xi_\text{norm}\\
	\xi_\text{$\theta$-tilt}
	}}
    \Big\lbrace
    \sum_{i=1}^5 \frac{[R_i- R^\text{th}_i]^2}
    {\sigma_{\text{stat},i}^2 + \sigma_{\text{sys},i}^2}
    + \xi_\text{norm}^2 + \xi^2_\text{$\theta$-tilt} \Big\rbrace \,,
    \\
    \label{eq:chi2mc}
    \chi^2_\text{cm}
    &= \frac{[R_\text{c}- R^\text{th}_\text{c}]^2}
    {\sigma_\text{stat,c}^2 + \sigma_\text{sys,c}^2}
    +
    \frac{[R_\text{m}- R^\text{th}_\text{m}]^2}
    {\sigma_\text{stat,m}^2 + \sigma_\text{sys,m}^2}
\end{align}
where $R^\text{th}_i = (1 + 0.2\, \xi_\text{norm}) \, (1 + 0.05\,
\langle \cos\theta \rangle_i \, \xi_\text{$\theta$-tilt})$ and we have
defined $R_\text{c}=R_{1}$ and 
\begin{equation}
    R_\text{m} =
    \frac{\displaystyle \sum_{i=2}^5 N_\mu^i(E_\mu^\text{fin} >
      E_\mu^\text{fin,min},\, \rho_\text{PREM})}
    {\displaystyle \sum_{i=2}^5 N_\mu^i(E_\mu^\text{fin} >
      E_\mu^\text{fin,min}, \rho_\text{hom})} \,,
\end{equation}
with $\sigma^2_\text{stat,m} = {\displaystyle \sum_{i=2}^5}
\sigma^2_{\text{stat},i}$ and $\sigma^2_\text{sys,m} = 4\times (1
\%)^2$.
In Eq.~\eqref{eq:chi2mc} $R^\text{th}_\text{c}$ and
$R^\text{th}_\text{m}$ are the corresponding theoretical predictions
including the normalization and tilt factors which minimize
$\chi^2_\text{nh}$. In the $\chi^2$ functions we have, conservatively,
not included the events in the most horizontal bin $\cos\theta >
-0.17$ where larger backgrounds from possible remaining
misreconstructed downgoing muons may be expected.  In choosing the
optimum energy threshold for this comparison, one has to take into
account that, as the energy increases, the Earth matter profile
becomes more evident but the statistics decreases and so does the
achievable precision.  For these simple observables, a compromise
sensitivity is achieved for $E_\mu^\text{fin,min}=10$~TeV. 

$\chi^2_\text{nh}$ quantifies the rejection power against the
hypothesis of a homogeneous Earth density, while $\chi^2_\text{cm}$
gives the sensitivity to the specific difference in density between
the core and the mantle.  We find that, if no deviation from the PREM
predictions is observed, in 10 years IceCube can reject the
homogeneity of the Earth with a $\chi^2_\text{nh} = 11.5$
($3.4\sigma$) which can reach $\chi^2_\text{nh} = 22$ ($4.7\sigma$) if
the theoretical and systematic uncertainties are reduced to be below
the statistical errors.  Correspondingly, the difference in density
between the core and mantle can be established with $\chi^2_\text{cm}
= \text{9--22}$ ($3\sigma$--$4.7\sigma$).

In summary, we conclude that IceCube will be able to measure the 
averaged core and mantle density with a significance that reaches,
conservatively, $3 \sigma$ in a decade. Our best guess is that a
result can be obtained at the $5 \sigma$ level. 

Work supported by Grants from U.S. NSF No.~OPP-0236449 and
PHY-0354776, and U.S.~DOE No.~DE-FG02-95ER40896, from Spanish MEC,
FPA-2004-00996, FPA-2006-01105 and FPA-2007-66665-C02-01 and from
Comunidad Aut\'onoma de Madrid P-ESP-00346.

\end{document}